\documentclass[a4paper,10pt,reqno]{amsart}
 \usepackage[a4paper,left=1in,right=1in]{geometry}
\usepackage{amssymb}
\usepackage{amsmath}
\usepackage{graphicx}
\usepackage{cite}
\usepackage{bm}
\usepackage[mathscr]{eucal}
\usepackage[usenames,dvipsnames]{color} 

\newcommand{\beq}{\begin{equation}}
\newcommand{\eeq}{\end{equation}}

\newcommand{\p}{\partial}

\begin{document}

\title{Asymmetric integrable quad-graph equations}

\author{Peter E. Hydon}
\address{Department of Mathematics\\ University of Surrey\\ Guildford\\ GU2 7XH\\ UK. }
\email{p.hydon@surrey.ac.uk}

\author{Claude-M. Viallet}
\address{LPTHE\\ UMR 7589 CNRS / Universit\'e Pierre et Marie Curie - Paris6\\ 4 Place Jussieu, Tour 24-25, 5\`eme Etage, Bo\^{\i}te 126 \\ F -- 75252 Paris, Cedex 05\\ FRANCE. }
\email{viallet@lpthe.jussieu.fr}

\date{\today}

\begin{abstract}
Integrable difference equations commonly have more low-order
conservation laws than occur for nonintegrable difference equations
of similar complexity. We use this empirical observation to sift a
large class of difference equations, in order to find candidates for
integrability. It turns out that all such candidates have an equivalent
affine form. These are tested by calculating their algebraic
entropy. In this way, we have found several types of integrable equations,
one of which seems to be entirely unrelated to any known discrete integrable
system. We also list all single-tile conservation laws for the integrable equations in the above class.
\end{abstract}

\maketitle

%

\section{Introduction}

A \textit{quad-graph equation} is a scalar difference equation for $u(k,l)$, where $(k,l)\in\mathbb{Z}^2$, which is of the form
\beq\label{qgr}
\mathcal{F}(k,l,u_{00},u_{10},u_{01},u_{11})=0.
\eeq
Here $u_{ij}$ denotes $u(k+i,l+j)$ and we assume that $\mathcal{F}$ depends on all four of these values. Various approaches have been used
to discover quad-graph equations that are integrable. Having developed the bilinear formalism for continuous integrable systems, Hirota discretized the bilinear operators for several known integrable systems, obtaining difference equations that had soliton solutions built-in \cite{hir1,hir2,hir3}. By contrast, Capel \textit{et al.} focused on discretizations of plane wave factors for the singular integral equations that are ubiquitous features of
continuous integrable systems \cite{qncv,npcq,nqc}. Whereas these approaches used discretizations of problems that were known to be integrable, Adler, Bobenko and Suris (ABS) dealt directly with quad-graph equations without reference to continuous systems. They obtained a classification of
all integrable quad-graph equations that are consistent on a cube (and thus admit a Lax pair), subject to certain nondegeneracy conditions \cite{ABS1,ABS2}. The idea that consistency on a cube is a sufficient condition for integrability was proposed independently by Nijhoff \cite{FLax} and Bobenko and Suris \cite{BSLax}.

To make further progress,
we adopt a different strategy. There is a systematic method for constructing conservation laws of difference equations; this has been used to
identify low-order conservation laws of many integrable quad-graph equations \cite{rh1,rh2}. From this work, we observe that integrable difference equations tend to have more low-order conservation laws than nonintegrable equations of similar complexity. Although this observation is purely empirical, we use it to sift a large class of quad-graph equations, in order to find equations that admit `extra' conservation laws. (This approach is dual to that of Levi and Yamilov, who recently obtained some necessary conditions for the existence of higher symmetries -- which again indicate integrability -- for certain types of quad-graph equations \cite{ly}). Having obtained a shortlist of possible candidates for integrability, we test their algebraic entropy.

Zero algebraic entropy is a signature of integrability \cite{FaVi93,HiVi98,Vi08}. This occurs for affine linear quad-graph equations when an arbitrary set of initial conditions produces polynomial growth in degree as one moves away from the initial points (see \S3 for details). Linear growth in degree
implies that the quad-graph equation is linearizable; all known integrable quad-graph equations that are not linearizable exhibit quadratic growth. The calculation of algebraic entropy is a diagnostic test, rather than a constructive method. For instance, Hietarinta discovered a quad-graph equation that is consistent on a cube, but does not appear in
the ABS list \cite{hiet}. A calculation of algebraic entropy showed that growth in degree for this quad-graph equation is linear; separately, Ramani \textit{et al.} found a clever linearization \cite{rjgt}.

In the next section, we determine conditions for the existence of extra conservation laws for a large class of quad-graph equations. Algebraic entropy is calculated in \S3, and we find that most of the sifted quad-graph equations exhibit quadratic growth in degree. For completeness, we list the conservation laws in \S4, before discussing our results and their consequences in \S5. 

\section{Classification of integrable cases via conservation laws}
\label{Class}
\noindent  In this section, we examine the conservation laws of equations of the form
\beq
u_{11}=\epsilon_1u_{00}+A(u_{10})-\epsilon_2A(u_{01}).
\label{First}
\eeq
Here each $\epsilon_i$ is either $1$ or $-1$, and $A$ is a nonlinear complex-valued function that is assumed to be `differentiable enough' (so that as many derivatives as needed are well-defined, at least locally). Equations in this class lack the $D_4$ symmetry of the ABS equations. The class includes some known integrable equations (such as the Lattice KdV equation) and is simple enough for a complete classification of integrable cases to be possible. Henceforth, we use $A_{ij}$ to denote $A(u_{ij})$. Conservation laws on a single tile satisfy the determining equation
\beq
F\big(k+1,\,l,\,u_{10},\,\omega\big)-F\big(k,\,l,\,u_{00},\,u_{01}\big)+G\big(k,\,l+1,\,u_{01},\,\omega\big)-G\big(k,\,l,\,u_{00},\,u_{10}\big)=0,
\label{Second}
\eeq
where $\omega$ denotes the right-hand side of (\ref{First}).  We solve (\ref{Second}) by deriving a sequence of its differential consequences, each of which eliminates at least one unknown function from the previous equation in the sequence (see \cite{rh1} for a fuller explanation).  This leads to an overdetermined system of functional--differential equations that can be solved completely.
Specifically, we apply the commuting differential operators
\[
 \mathcal{L}_1=\p_{10}-\epsilon_1A'_{10}\p_{00},\qquad \mathcal{L}_2=\p_{01}+\epsilon_1\epsilon_2A'_{01}\p_{00},
\]
(where $\p_{ij}$ denotes $\p/\p u_{ij}$) to obtain
\beq
\big(\epsilon_2A'_{10}A'_{01}\p_{00}+\epsilon_1A'_{10}\p_{01}\big)\p_{00}F\big(k,\,l,\,u_{00},\,u_{01}\big)
+\big(\epsilon_2A'_{10}A'_{01}\p_{00}-\epsilon_1\epsilon_2A'_{01}\p_{10}\big)\p_{00}G\big(k,\,l,\,u_{00},\,u_{10}\big)=0.
\label{Third}
\eeq
Dividing by $\epsilon_2A'_{10}A'_{01}$, then differentiating with respect to $u_{01}$ yields the partial differential equation
\beq
\p_{01}\Big(\p_{00}+\frac{\epsilon_1\epsilon_2}{A'_{01}}\p_{01}\Big)\p_{00}F\big(k,\,l,\,u_{00},\,u_{01}\big)=0, \label{Fourth}
\eeq
whose general solution is
\beq
F\big(k,\,l,\,u_{00},\,u_{01}\big)=f_1\big(k,\,l,\,\epsilon_1u_{00}-\epsilon_2A_{01}\big)+f_2\big(k,\,l,\,u_{00}\big)+f_3\big(k,\,l,\,u_{01}\big). \label{Fifth}
\eeq
Without loss of generality, set $f_3=0$ (absorbing the resulting trivial conservation law into $f_2$ and $G$).  Then (\ref{Third}) amounts to
\[
\Big(\p_{00}-\frac{\epsilon_1}{A'_{10}}\p_{10}\Big)\p_{00}G\big(k,\,l,\,u_{00},\,u_{10}\big)=-\p^2_{00}f_2\big(k,\,l,\,u_{00}\big),
\]
whose general solution is
\beq
G\big(k,\,l,\,u_{00},\,u_{10}\big)=g_1\big(k,\,l,\,\epsilon_1u_{00}+A_{10}\big)+g_2\big(k,\,l,\,u_{10}\big)-f_2\big(k,\,l,\,u_{00}\big). \label{Sixth}
\eeq
At this stage, it is convenient to substitute (\ref{Fifth}) and (\ref{Sixth}) into the determining equation (\ref{Second}), using the difference equation (\ref{First}) to eliminate $u_{00}$.  This puts the determining equation in the form
\begin{eqnarray} f_1\Big(k+1,\,l,\,\epsilon_1u_{10}-\epsilon_2A_{11}\Big)-f_1\Big(k,\,l,\,u_{11}-A_{10}\Big)+f_2\big(k+1,\,l,\,u_{10}\big)
-f_2\big(k,\,l+1,\,u_{01}\big)\qquad&&\nonumber\\
+\;g_1\Big(k,\,l+1,\,\epsilon_1u_{01}+A_{11}\Big)-g_1\Big(k,\,l,\,u_{11}+\epsilon_2A_{01}\Big) +g_2\big(k,\,l+1,\,u_{11}\big)-g_2\big(k,\,l,\,u_{10}\big)=0.&&\label{Seven}
\end{eqnarray}
Applying $\p_{01}\p_{11}$ to (\ref{Seven}), we obtain
\beq
\epsilon_1A'_{11}g''_1\Big(k,\,l+1,\,\epsilon_1u_{01}+A_{11}\Big)=\epsilon_2A'_{01}g''_1\Big(k,\,l,\,u_{11}+\epsilon_2A_{01}\Big),
\label{Eight}
\eeq
where $g''_1$ is the second derivative of $g$ with respect to its third argument.  This condition holds trivially if $g_1$ is linear in the third argument, which leads to two `universal' conservation laws for which $F$ and $G$ are each linear in $A_{ij}$.  These are
\begin{align}
&F_1\big(k,\,l,\,u_{00},\,u_{01}\big)=\big(\sqrt{\epsilon_1\epsilon_2}\big)^{k+l-1}\epsilon_2^l\big(\epsilon_2u_{00}-\sqrt{\epsilon_1\epsilon_2}A_{01}\big),&\nonumber\\
&G_1\big(k,\,l,\,u_{00},\,u_{10}\big)=\big(\sqrt{\epsilon_1\epsilon_2}\big)^{k+l}\epsilon_2^l\big(\epsilon_2u_{10}+A_{00}\big),& \label{Nine}
\end{align}
and
\begin{align}
&F_2\big(k,\,l,\,u_{00},\,u_{01}\big)=\big(-\sqrt{\epsilon_1\epsilon_2}\big)^{k+l-1}\epsilon_2^l\big(\epsilon_2u_{00}+\sqrt{\epsilon_1\epsilon_2}A_{01}\big),&\nonumber\\
&G_2\big(k,\,l,\,u_{00},\,u_{10}\big)=\big(-\sqrt{\epsilon_1\epsilon_2}\big)^{k+l}\epsilon_2^l\big(\epsilon_2u_{10}+A_{00}\big),&  \label{Ten}
\end{align}

In order to find all functions $A$ for which there are additional conservation laws on a tile, we now restrict attention to the case $g''_1\neq 0$.\footnote{If $g_1''=0$ but $f_1''\neq 0$, similar calculations lead to precisely the same classifying equation (\ref{Thirteen}), so nothing is lost by this assumption.} Dividing (\ref{Eight}) by $A'_{01}$, then applying the operator $\p_{01}-\epsilon_2A'_{01}\p_{11}$ and rearranging the result, we obtain
\[
 \frac{g'''_1\big(k,\,l+1,\epsilon_1u_{01}+A_{11}\big)}{g''_1\big(k,\,l+1,\,\epsilon_1u_{01}+A_{11}\big)}=\frac{A'_{11}A''_{01}+\epsilon_2\big(A'_{01}\big)^2A''_{11}}{A'_{01}A'_{11}\big(\epsilon_1-\epsilon_2A'_{01}A'_{11}\big)}.
\]
It is convenient to write $A'_{ij}=B\big(A_{ij}\big)\equiv B_{ij}$, so that $A''_{ij}=B_{ij}B'_{ij}$ (which is nonzero, as $A_{ij}$ is a nonlinear function of $u_{ij}$) and $A'''_{ij}=\big(B_{ij}\big)^2B''_{ij}+B_{ij}\big(B'_{ij}\big)^2$. Then
\beq
\frac{g'''_1\big(k,\,l+1,\epsilon_1u_{01}+A_{11}\big)}{g''_1\big(k,\,l+1,\,\epsilon_1u_{01}+A_{11}\big)}=\frac{B'_{01}+\epsilon_2B_{01}B'_{11}}{\epsilon_1-\epsilon_2B_{01}B_{11}}.
\label{Eleven}
\eeq
Applying the operator
\[
 \p_{01}-\frac{\epsilon_1}{A'_{11}}\,\p_{11}=B_{01}\frac{\p}{\p A_{01}}-\epsilon_1\frac{\p}{\p A_{11}}
\]
to (\ref{Eleven}) gives (after simplification)
\beq
\big(1-\epsilon_1\epsilon_2B_{01}B_{11}\big)\big(B''_{01}-\epsilon_1\epsilon_2B''_{11}\big)-B_{01}\big(B'_{11}\big)^2+\epsilon_1\epsilon_2B_{11}\big(B'_{01}\big)^2=0.
\label{Twelve}
\eeq
This is the classifying equation that yields all functions $A$ for which there exist conservation laws other than (\ref{Nine}) and (\ref{Ten}). As (\ref{Twelve}) stands, the functions $B_{01}$ and $B_{11}$ are thoroughly entwined, but this can be resolved by one further differentiation, which yields the necessary condition
\[
 \big(B'''_{ij}/B'_{ij}\big)'=0.
\]
A simple calculation shows that $\big(B'_{ij}\big)^2$ is a nonzero quadratic function of $B_{ij}$; substituting this into (\ref{Twelve})
 and solving the resulting conditions gives
\beq
\big(B'_{ij}\big)^2=c_1^2\big(B^2_{ij}+1\big)+(1+\epsilon_1\epsilon_2)c_2B_{ij},\qquad c_1,c_2\in\mathbb{C}.
\label{Thirteen}
\eeq
(Here and henceforth, arbitrary constants are denoted $c$ or $c_i$.) This splits into four cases, as follows.
\vspace*{0.5cm}

\noindent \textbf{Case I}:\; $c_1=0$.
\vspace*{0.2cm}

In this case, we require $\epsilon_2=\epsilon_1$ and $c_2\neq 0$, in order that $B'_{ij}$ is nonzero.  Then
\[
 A'_{ij}=B_{ij}=\frac{c_2}{2}\big(A_{ij}+c_3\big)^2,
\]
so
\beq
A_{ij}=\frac{c}{(c_4-u_{ij})} - c_3,\qquad\text{where}\quad c=2/c_2\neq 0.
\label{Fourteen}
\eeq
Then the solution of (\ref{Eleven}), after absorbing the linear terms into $f_2$ and $g_2$, is
\beq
g_1\big(k,\,l+1,\,\epsilon_1u_{01}+A_{11}\big)=c_5\ln\big(\epsilon_1c_4-c_3-(\epsilon_1u_{01}+A_{11})\big),\qquad c_5\neq 0.
\label{Fifteen}
\eeq
This satisfies (\ref{Eight}) if $\epsilon_1=\epsilon_2=1$, but if $\epsilon_1=\epsilon_2=-1$ then (\ref{Eight}) gives the further constraint $c_3=c_4$. So this case leads to two possible equations, namely
\beq
u_{11}=u_{00}- c\Big(\frac{1}{u_{10}}-\frac{1}{u_{01}}\Big),
\label{Sixteen}
\eeq
and
\beq
u_{11}=-u_{00}- c\Big(\frac{1}{u_{10}}+\frac{1}{u_{01}}\Big),
\label{Seventeen}
\eeq
where the constant $c_4$ has been absorbed into $u_{ij}$.  The point transformation $u_{00}\mapsto(-1)^ku_{00}$ maps (\ref{Seventeen}) into (\ref{Sixteen}), which is the lattice KdV equation (simplified slightly from the form stated in \cite{grp}). By solving (\ref{Seven}) for the remaining unknown functions, we obtain five conservation laws for (\ref{Sixteen}), which are listed later (after all integrable quad-graph equations of the form (\ref{First}) have been identified). The corresponding conservation laws for (\ref{Seventeen}) follow from the above transformation.
\vspace*{0.8cm}\

\noindent \textbf{Case II}:\; $c_1\neq 0,\,\epsilon_2=\epsilon_1,\,c_2^2=c_1^4$.
\vspace*{0.2cm}

In this case set $c_2=\epsilon_3c_1^2$, where $\epsilon_3=\pm 1$.  Then the general solution of (\ref{Thirteen}) leads to the result
\beq
 e^{c_1A_{ij}+c_3}=\frac{\epsilon_3}{1-z_{ij}^{\epsilon_3}},\label{twentyone}
\eeq
where the notation $z_{ij}=e^{c_1 u_{ij}+c_4}$ is used henceforth. Then (\ref{Eleven}) amounts to
\beq
\frac{g_1'''\big(k,\,l+1,\,\epsilon_1u_{01}+A_{11}\big)}{g_1''\big(k,\,l+1,\,
\epsilon_1u_{01}+A_{11}\big)}=\frac{c_1\big[\epsilon_1
+z_{01}^{\epsilon_3}e^{c_1A_{11}+c_3}\big]}{\epsilon_3-z_{01}^{\epsilon_3}e^{c_1A_{11}+c_3}}\,.\label{Twentytwo}
\eeq
If $\epsilon_3=\epsilon_1$, the general solution of (\ref{Twentytwo}) is
\[
g_1''\big(k,\,l+1,\,\epsilon_1u_{01}+A_{11}\big)=
\frac{a(k,l+1)\exp\big\{c_1(\epsilon_1u_{01}+A_{11})+c_3+\epsilon_1c_4\}}{\big[1-\epsilon_1\exp\big\{c_1(\epsilon_1u_{01}+A_{11})+c_3+\epsilon_1c_4\}\big]^2}\,,
\]
where $a(k,l)$ is an arbitrary nonzero function. So when $\epsilon_3=\epsilon_1=1$, the condition (\ref{Eight}) gives only $a(k,l)=\alpha(k)$, whereas when $\epsilon_3=\epsilon_1=-1$ it also gives the constraint $e^{2(c_3-c_4)}=1$. Writing (\ref{First}) in terms of $z_{ij}$, we obtain
\beq
\frac{z_{11}}{z_{00}}=\frac{z_{01}-1}{z_{10}-1}\,,\qquad\qquad (\text{when } \epsilon_3=\epsilon_1=1),\label{Twentyfour}
\eeq
and
\beq
z_{00}z_{11}=\frac{z_{10}z_{01}}{(z_{10}-1)(z_{01}-1)}\,,\qquad\qquad (\text{when }\epsilon_3=\epsilon_1=-1).\label{Twentyfive}
\eeq
In \cite{rgsw}, Ramani \textit{et al.} show that (\ref{Twentyfour}) is equivalent (under a point transformation) to the `discrete Lotka--Volterra equation of type I' that was discovered by Hirota and Tsujimoto \cite{ht95}. Levi and Yamilov recently found higher symmetries, a Lax pair and two conservation laws for a variant of this equation \cite{ly}.

When $\epsilon_3=-\epsilon_1$, equation (\ref{Twentytwo}) yields
\[
g_1''(k,\,l+1,\,\epsilon_1u_{01}+A_{11})=a(k,l+1)\exp\big\{-c_1(\epsilon_1u_{01}+A_{11})-c_3-\epsilon_1c_4\big\}
\]
Equation (\ref{Eight}) produces the constraint  $a(k,l)=\alpha(k)$ when $\epsilon_3=-\epsilon_1=-1$, but when
$\epsilon_3=-\epsilon_1=1$ it gives $a(k,l)=0$. So we obtain only one further equation, namely
\beq
\frac{z_{11}}{z_{00}}=\frac{z_{10}(z_{01}-1)}{z_{01}(z_{10}-1)},\qquad\qquad (\text{when }\epsilon_3=-\epsilon_1=-1).
\label{tseven}
\eeq
Note that (\ref{Twentyfour}), (\ref{Twentyfive}) and (\ref{tseven}) are affine linear in each $z_{ij}$. \vspace*{0.8cm}

\noindent\textbf{Case III}:\;$c_1\neq 0,\,\epsilon_2=\epsilon_1,\,c_2^2\neq c_1^4$.
\vspace*{0.2cm}

\noindent In this case
\[
B_{ij}=\bar{c}_2\,\text{sinh}\,\big(c_1 A_{ij}+c_3\big)-\tilde c_2
\]
where $\tilde c_2=c_2/c_1^2$ and $\bar{c}_2^2=1-\tilde{c}_2^2 \neq 0$. Then the general solution of $A'_{ij}=B_{ij}$ is
\[
 e^{c_1A_{ij}+c_3} = \frac{1+\tilde c_2+(1-\tilde c_2)z_{ij}}{\bar{c}_2(1-z_{ij})}\,,
\]
and therefore (\ref{Eleven}) amounts to
\[
\frac{g_1'''\big(k,\,l+1,\,\epsilon_1u_{01}+A_{11}\big)}{g_1''\big(k,\,l+1,\,\epsilon_1u_{01}+A_{11}\big)}=\frac{c_1\Big[1+\epsilon_1\tilde c_2+\epsilon_1\bar{c}_2\exp\big\{c_1(\epsilon_1u_{01}+A_{11})+c_3+\epsilon_1c_4\big\}\Big]}
{\Big[1+\epsilon_1\tilde c_2-\epsilon_1\bar{c}_2\exp\big\{c_1(\epsilon_1u_{01}+A_{11})+c_3+\epsilon_1c_4\big\}\Big]}\,.
\]
Hence
\[
 g_1''(k,\,l+1,\,\epsilon_1u_{01}+A_{11})=\frac{a(k,l+1)\exp\big\{c_1(\epsilon_1u_{01}+A_{11})+c_3+\epsilon_1c_4\big\}}{\Big[1+\epsilon_1\tilde c_2-\epsilon_1\bar{c}_2\exp\big\{c_1(\epsilon_1u_{01}+A_{11})+c_3+\epsilon_1c_4\big\}\Big]^2}\,,
\]
and so (\ref{Eight}) produces the constraint $a(k,l)=\alpha(k)$ when $\epsilon_1=1$. The resulting difference equation is
\beq
\frac{z_{11}}{z_{00}}=\frac{(z_{10}+c)(z_{01}-1)}{(z_{01}+c)(z_{10}-1)}\,,\qquad\qquad c\notin\{-1,0\},
\label{teight}
\eeq
where $c=(1+\tilde c_2)/(1-\tilde c_2)$. This is equivalent under a point transformation to the
lattice MKdV equation\footnote{We thank Frank Nijhoff and Kenichi Maruno for alerting us to this.} (see \cite{ht,mkno,nah}); in particular, a Lax pair for this equation is given in \cite{nah}.

When $\epsilon_1=-1$, the condition (\ref{Eight}) yields $a(k,l)=\alpha(k)$, together with $e^{2(c_4-c_3)}=c$. This leads to the difference equation
\beq
z_{00}z_{11}=\frac{(z_{10}+c)(z_{01}+c)}{(z_{10}-1)(z_{01}-1)}\,,\qquad\qquad c\notin\{-1,0\}.
\label{tnine}
\eeq
The point transformation
\beq
z_{ij}\mapsto(-c)^{k+i}z_{ij}^{(-1)^{k+i}} \label{pt1}
\eeq
maps (\ref{tnine}) to (\ref{teight}). Note that when $c=0$, (\ref{teight}) reduces to (\ref{tseven}). Furthermore, (\ref{Twentyfour}) is the limit of (\ref{teight}) as $c\rightarrow\infty$ with $z_{ij}$ fixed. So (\ref{Twentyfour}) and (\ref{tseven}) are each
singular limits of the lattice MKdV equation. Moreover, the point transformation
\beq
\label{pt2}
z_{ij}\mapsto1/z_{ji}
\eeq
maps (\ref{tseven}) to the Lotka-Volterra type equation (\ref{Twentyfour}).\footnote{We are grateful to an anonymous referee for this observation.}
\vspace*{0.8cm}

\noindent\textbf{Case IV}:\; $c_1\neq 0,\,\epsilon_2=-\epsilon_1$.
\vspace*{0.2cm}

\noindent This is similar to Case III; the solution of (\ref{Thirteen}) is
\[
 B_{ij}=\text{sinh}\,(c_1A_{ij}+c_3).
\]
Therefore
\beq
e^{c_1A_{ij}+c_3}=\frac{1+z_{ij}}{1-z_{ij}}\,,
\label{thirty}
\eeq
and so
\[
 \frac{g_1'''\big(k,\,l+1,\,\epsilon_1u_{01}+A_{11}\big)}{g_1''\big(k,\,l+1,\,\epsilon_1u_{01}+A_{11}\big)}=\frac{c_1\big[1-\epsilon_1\exp\big\{c_1(\epsilon_1u_{01}+A_{11})+c_3+\epsilon_1c_4\big\}\big]}
{\big[1+\epsilon_1\exp\big\{c_1(\epsilon_1u_{01}+A_{11})+c_3+\epsilon_1c_4\big\}\big]}\,.
\]
Hence
\[
g_1''\big(k,\,l+1,\,\epsilon_1u_{01}+A_{11}\big)=\frac{a(k,l+1)\exp\big\{c_1(\epsilon_1u_{01}+A_{11})+c_3+\epsilon_1c_4\big\}}{\Big[1+\epsilon_1\exp\big\{c_1(\epsilon_1u_{01}+A_{11})+c_3+\epsilon_1c_4\big\}\Big]^2}\,.
\]
When $\epsilon_1=1$, (\ref{Eight}) gives the constraints $a(k,l)=\alpha(k)$ and $e^{2c_3}=-1$, and the resulting difference equation is
\beq
\frac{z_{11}}{z_{00}}=-\,\frac{(z_{10}+1)(z_{01}+1)}{(z_{10}-1)(z_{01}-1)}\,.
\label{thone}
\eeq
When $\epsilon_1=-1$, we obtain similarly $a(k,l)=\alpha(k),\,e^{2c_4}=-1$, which leads to
\beq
z_{00}z_{11}=-\,\frac{(z_{10}+1)(z_{01}-1)}{(z_{10}-1)(z_{01}+1)}\,.
\label{thtwo}
\eeq

Once again, the process has produced affine linear equations. It turns out that (\ref{thone}) can be mapped to (\ref{thtwo}) by the point transformation (\ref{pt1}) with $c=1$.

\section{Algebraic entropy}

To test the integrability of the previous lattice maps, we evaluate
their algebraic entropy~\cite{BeVi99,TrGrRa01,Vi06}. The system has an
infinite dimensional space of initial conditions. We choose initial
conditions on a diagonal regular staircase, which is shown in Figure 1.
\begin{eqnarray}
\Delta=\big\{u_{nm}:\ n+m \in\{0,1\}\big\}.
\end{eqnarray}
This defines a forward evolution towards the
upper right corner of the lattice, and a backward evolution towards
the lower left corner.

\begin{figure}[htp]
\centering
\includegraphics{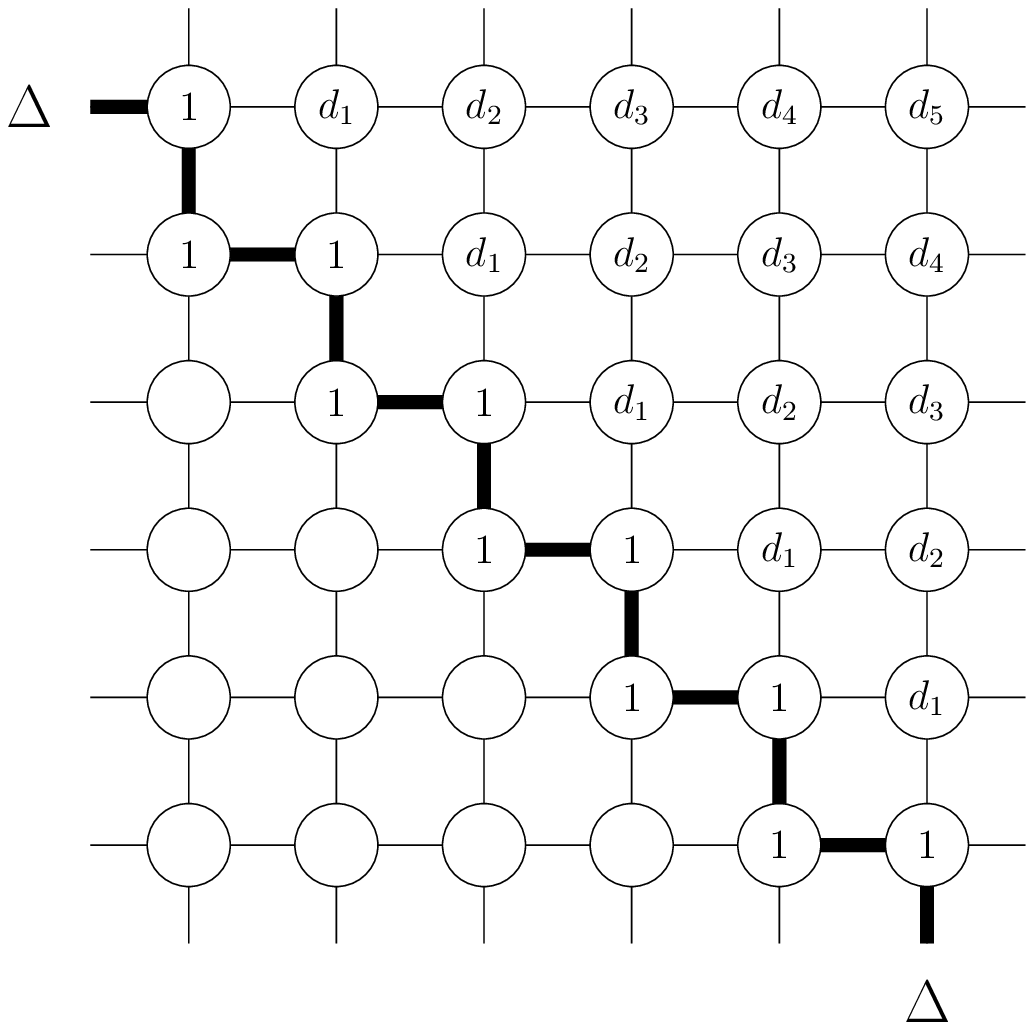}
\caption{The distribution of degrees over the lattice.}\label{fig1}
\end{figure}

The method is to let the system evolve, calculating $u_{nm}$ away from
the diagonal by using (recursively) the defining relation on an
elementary tile of the lattice. Each $u_{nm}$ is a rational polynomial in
terms of the initial conditions; the degree of the denominator is
evaluated. The space of initial conditions is infinite-dimensional
but, for any quad-graph equation, we need to specify only $2k+1$ initial
conditions to evaluate $k$ iterates. This gives a sequence of degrees
$\{ d_n \}$, as shown in Figure 1. The growth of that sequence gives
the entropy
\begin{equation}
\epsilon = \lim_{n\rightarrow\infty}\frac{1}{n}\;  \ln(d_n) .
\end{equation}
Vanishing of the entropy is the hallmark of integrability~\cite{FaVi93,HiVi98,Vi08}.

Although we are able to calculate only a limited number of terms of
the sequence, it is possible to infer the exact value of the
entropy. The reason is the existence of a finite recurrence with
integer coefficients that is satisfied by the sequence of degrees. The most
efficient way to find this recurrence is to fit the sequence with a
Pad\'e approximant. The existence of the recurrence on the degrees
ensures that the generating function for the sequence of degrees is a
rational fraction.

Table 1 gives the sequences of degrees and the
corresponding entropy for the various quad-graph equations in \S\ref{Class}. For
comparison, we also include a nonintegrable equation that is only slightly different to (\ref{tnine}),
namely
\beq
z_{00}z_{11}=\frac{(z_{10}+2)(z_{01}+2)}{2(z_{10}-1)(z_{01}-1)}.
\label{tnineb}
\eeq


\begin{table}[hp]
\begin{tabular}{||c|c|c|c||}
\hline
& & & \\
   Equation & Sequence & $\{ d_n \}$ & $\epsilon$    \\
& & & \\
\hline
& & & \\
   (\ref{Sixteen}) &  $1, 3, 7, 13, 21, 31, 43, 57, \dots$ &  $1+n+n^2$ & 0    \\
& & & \\
   (\ref{Twentyfour}) & $1,2, 4, 7, 11,16, 22, 29,\dots$ & $ 1+(n^2+n)/2$  & 0    \\
& & & \\
   (\ref{Twentyfive}) & $1, 3, 6, 10, 14, 18, 22, 26,\dots$  & $4n-2,\;(n\geq 2)$  & 0  \\
& & & \\
   (\ref{tseven}) & $ 1, 3, 6, 11, 18, 27, 38, 51 ,\dots$ & $n^2+2, (n \geq 1)$  & 0   \\
& & & \\
   (\ref{teight}) & $ 1, 3, 7, 13, 21, 31, 43, 57 ,\dots$  &  $1+n+n^2$  & 0    \\
& & & \\
   (\ref{thtwo}) & $1,3,7,13,21,31,43,57,\dots$  &  $1+n+n^2$  & 0     \\
& & & \\
   (\ref{tnineb}) & $1,3,7,17,41,99,239,577,\dots$ & $((1+\sqrt{2})^{n+1} + (1-\sqrt{2})^{n+1})/2$ & $\ln(1+\sqrt{2})$   \\[3pt]
\hline
\end{tabular}
\medskip
\caption{The sequence of degrees for each equation.}\label{tab1}
\end{table}

\bigskip

Equation (\ref{Twentyfive}) has linear growth of the degree, which
indicates that this equation is linearizable. This result is confirmed
by the existence of an infinite family of conservation laws on a
single tile (see \S4). However, we have not been able to discover a linearizing
transformation. The point transformation $z_{ij}\mapsto 1/z_{ij}$ simplifies
(\ref{Twentyfive}) to 
\beq
z_{00}z_{11}=(z_{10}-1)(z_{01}-1),
\eeq
but this is no more tractable than the original equation.

All other cases that have more than two conservation laws
exhibit quadratic growth of the degree, and therefore
are claimed to be integrable, but not linearizable.  This raises the
question of whether any of the new integrable quad-graph equations can be
mapped to any known equation. This will be discussed in \S5.

\section{The conservation laws}

\noindent Although the lattice KdV equation (\ref{Sixteen}), the lattice MKdV equation (\ref{teight}) and
the discrete Lotka-Volterra equation (\ref{Twentyfour}) are not
new, their conservation laws (on a single tile) have not previously been
listed.  Throughout this section, the universal conservation laws (\ref{Nine}) and
(\ref{Ten}) span the first two conservation laws in each list (up to
the addition of trivial conservation laws).  So the `extra'
conservation laws are $(F_i,G_i)$, where $i\geq3$; here $F_i$ denotes
$F_i(k,\,l,\,z_{00},\,z_{01})$ and $G_i$ denotes
$G_i(k,\,l,\,z_{00},\,z_{10})$. Lattice KdV has the following conservation laws.
\begin{flalign*}
F_1&=u_{00}+c/u_{01},\hspace*{3cm} &G_1&=u_{10}-c/u_{10};\\[4pt]
F_2&=(-1)^{k+l}\big(u_{00}+c/u_{01}\big),&G_2&=(-1)^{k+l+1}\big(u_{10}+c/u_{10}\big);\\[4pt]
F_3&=\ln\big(u_{00}+c/u_{01}\big),&G_3&=\ln(u_{10});\\[4pt]
F_4&=\ln(u_{00}),&G_4&=\ln\big(u_{10}-c/u_{00}\big);\\[4pt]
F_5&=-k\ln(u_{00})+l\ln\big(u_{00}+c/u_{01}\big),\qquad&G_5&=-k\ln\big(u_{10}-c/u_{00}\big)+(l-1)\ln(u_{10}).
\end{flalign*}

\noindent For lattice MKdV, the single-tile conservation laws are:
\begin{flalign*}
 F_1&=\ln\bigg(\frac{z_{00}(z_{01}-1)}{z_{01}+c}\bigg), &G_1=&\ln\bigg(\frac{z_{10}(z_{10}+c)}{z_{10}-1}\bigg);&\\[4pt]
F_2&=(-1)^{k+l}\ln\bigg(\frac{z_{01}+c}{z_{00}(z_{01}-1)}\bigg), &G_2=&(-1)^{k+l}\ln\bigg(\frac{z_{10}(z_{10}-1)}{z_{10}+c}\bigg);&\\[4pt]
F_3&=\ln\bigg(\frac{z_{00}z_{01}-z_{00}-z_{01}-c}{z_{01}+c}\bigg), &G_3=&\ln(z_{10}+c);&\\[4pt]
F_4&=\ln\bigg(\frac{z_{00}+c}{z_{00}}\bigg), &G_4=&\ln\bigg(\frac{z_{00}z_{10}+cz_{00}+cz_{10}-c}{z_{10}(z_{00}+c)}\bigg);&\\
\end{flalign*}
\vspace*{-1cm}
\begin{flalign*}
&F_5=k\ln\bigg(\frac{z_{00}(z_{01}+c)}{(z_{01}-1)(z_{00}+c)^2}\bigg)+l\ln\bigg(\frac{(z_{00}z_{01}-z_{00}-z_{01}-c)^2}{z_{00}(z_{01}+c)(z_{01}-1)}\bigg),&\\[4pt]
&\phantom{0}\qquad G_5=k\ln\bigg(\frac{z_{10}(z_{10}-1)(z_{00}+c)^2}{(z_{10}+c)(z_{00}z_{10}+cz_{00}+cz_{10}-c)^2}\bigg)+l\ln\bigg(\frac{(z_{10}+c)(z_{10}-1)}{z_{10}}\bigg)+\ln\bigg(\frac{z_{10}}{(z_{10}+c)^2}\bigg).&
\end{flalign*}


\noindent The single-tile conservation laws for the discrete Lotka-Volterra equation (\ref{Twentyfour}) are
\begin{flalign*}
 F_1&=\ln\big(z_{00}(z_{01}-1)\big), &G_1=&\ln\bigg(\frac{z_{10}}{z_{10}-1}\bigg);&\\[4pt]
F_2&=(-1)^{k+l}\ln\big(z_{00}(z_{01}-1)\big), &G_2=&(-1)^{k+l+1}\ln\big(z_{10}(z_{10}-1)\big);&\\[4pt]
F_3&=z_{00}(1-z_{01}), &G_3=&z_{10};&\\[4pt]
F_4&=\ln(z_{00}), &G_4=&\ln\bigg(\frac{z_{10}}{z_{00}+z_{10}-1}\bigg);&\\[4pt]
F_5&=k\ln\bigg(\frac{z_{00}}{z_{01}-1}\bigg)+l\ln\big(z_{00}(z_{01}-1)\big),\qquad &G_5=&k\ln\bigg(\frac{z_{10}(z_{10}-1)}{(z_{00}+z_{10}-1)^2}\bigg)+l\ln\bigg(\frac{z_{10}-1}{z_{10}}\bigg)+\ln(z_{10}).&
\end{flalign*}

\vspace*{-0.1cm}
\noindent Levi and Yamilov \cite{ly} recently derived an alternative form of (\ref{Twentyfour}) and listed two of its conservation laws, which are equivalent to $(F_1,G_1)$ and $(F_4,G_4)$.

We now list the conservation laws corresponding to the remaining affine linear
quad-graph equations that we have derived. Each of our equations that is not equivalent to Lattice KdV, Lattice MKdV or the Lotka-Volterra
type equation is equivalent to either the linearizable equation
(\ref{Twentyfive}) or the new equation (\ref{thtwo}).
\medskip

\noindent\textbf{Equation (\ref{Twentyfive})}
\vspace*{0.2cm}

\noindent This equation has an infinite set of conservation laws, which depend upon two arbitrary functions $\alpha,\beta$:
\[
 F_{\alpha} =\alpha(l+1)\ln\bigg(\frac{z_{00}z_{01}-z_{00}-z_{01}}{z_{00}(z_{01}-1)}\bigg) +\alpha(l)\ln\bigg(\frac{z_{00}z_{01}-z_{00}-z_{01}}{z_{01}}\bigg),\qquad\qquad\  G_{\alpha}=\alpha(l)\ln(1-z_{10});
\]
\[
F_{\beta}=\beta(k)\ln(1-z_{01}),\qquad\qquad G_{\beta}=\beta(k+1)\ln\bigg(\frac{z_{00}z_{10}-z_{00}-z_{10}}{z_{00}(z_{10}-1)}\bigg)+\beta(k)\ln\bigg(\frac{z_{00}z_{10}-z_{00}-z_{10}}{z_{10}}\bigg).
\]

\noindent This is a further indicator that, unlike the other quad-graph equations in our class, (\ref{Twentyfive}) is linearizable.
\vspace*{0.5cm}

%

\noindent\textbf{Equation (\ref{thtwo})}
\begin{flalign*}
 F_1&= \big(-1\big)^{(k+l)(k+l-1)/2}\ln\Big(\tfrac{z_{00}(z_{01}+1)}{z_{01}-1}\Big),\quad &G_1=&\cos\Big(\tfrac{(k+l)\pi}{2}\Big)\ln\Big(\tfrac{z_{10}(1-z_{10})}{z_{10}+1}\Big)+\sin\Big(\tfrac{(k+l)\pi}{2}\Big)\ln\Big(\tfrac{z_{10}-1}{z_{10}(z_{10}+1)}\Big);&\\[5pt]
 F_2&= \big(-1\big)^{(k+l)(k+l+1)/2}\ln\Big(\tfrac{z_{00}(z_{01}+1)}{z_{01}-1}\Big),\quad &G_2=&-\sin\Big(\tfrac{(k+l)\pi}{2}\Big)\ln\Big(\tfrac{z_{10}(1-z_{10})}{z_{10}+1}\Big)+\cos\Big(\tfrac{(k+l)\pi}{2}\Big)\ln\Big(\tfrac{z_{10}-1}{z_{10}(z_{10}+1)}\Big);&\\[5pt]
F_3&=\ln\bigg(\frac{(z_{00}+1)^2(z_{01}-1)}{z_{00}(z_{01}+1)}\bigg),\quad\; &G_3=&\ln\bigg(\frac{(-1)^l(z_{00}z_{10}-z_{00}+z_{10}+1)^2(z_{10}+1)}{z_{10}(z_{00}+1)^2(z_{10}-1)}\bigg);&\\[5pt]
F_4&=\ln\bigg(\frac{(z_{00}z_{01}+z_{00}-z_{01}+1)^2}{z_{00}(z_{01}+1)(z_{01}-1)}\bigg),\quad &G_4=&\ln\bigg(\frac{(-1)^l(z_{10}+1)(z_{10}-1)}{z_{10}}\bigg);&\\
\end{flalign*}
\vspace*{-0.8cm}
\begin{flalign*}
&F_5=k\ln\bigg(\frac{z_{00}(z_{01}+1)}{(z_{00}+1)^2(z_{01}-1)}\bigg)
+l\ln\bigg(\frac{(-1)^k(z_{00}z_{01}+z_{00}-z_{01}+1)^2}{z_{00}(z_{01}+1)(z_{01}-1)}\bigg),&\\[5pt]
&\phantom{0}\qquad G_5=k\ln\bigg(\frac{(-1)^lz_{10}(z_{00}+1)^2(z_{10}-1)}{(z_{10}+1)(z_{00}z_{10}-z_{00}+z_{10}+1)^2}\bigg)
+l\ln\bigg(\frac{(z_{10}+1)(z_{10}-1)}{z_{10}}\bigg)+\ln\bigg(\frac{z_{10}}{(z_{10}+1)^2}\bigg).&
\end{flalign*}

\section{Comments}
\label{Comments}

Remarkably, although the original Ansatz (\ref{First})
contained an arbitrary function $A$, each of the equations that we have found by sifting
can be written in affine form, using a simple change of dependent variable.
This has made the entropy calculation possible, because it gives
rational evolution.

It is natural to ask at this point how the equations that we have derived compare to
the known affine linear quad-graph equations. We have already seen that in most cases,
our Ansatz yields an equation that is equivalent under a point transformation
to a known equation. Therefore it is important to characterize this equivalence, which can be done
using the appoach introduced in \cite{ABS2}. Any affine linear quad-graph equation can be written in
polynomial form:
\begin{equation*}
Q(v_1, v_2, v_3, v_4) = 0,
\end{equation*}
where $v_i,\; i=1\dots 4$, are the values (of $u_{ij}$ or $z_{ij}$ as appropriate) at the four corners.  For
any choice of a pair of indices $1 \leq i < j \leq 4$, define $h_{ij}$
by
\begin{equation}
h_{ij}(v_k,v_l) = \partial_{v_i} Q
\cdot \partial_{v_j} Q - Q \cdot \partial_{v_i} \partial_{v_j} Q, \qquad
i \neq j \neq k \neq l
\end{equation}
It is then possible to associate to each of the four corners a polynomial
\begin{equation}
r_k(v_k) = (\partial_{v_l} h_{ij})^2- 2 \;h_{ij} \; (\partial_{v_l}^2 h_{ij}).
\end{equation}
These polynomials play a central role in the classification of~\cite{ABS2}, because (after a M\"{o}bius
tranformation, if necessary), they can take one of six canonical forms, according to their root distribution.

For example, the lattice MKdV equation (\ref{teight}) yields (adjusting the notation for clarity)
\begin{flalign*}
h_{z_{00} z_{01}} & =  (1+c) \; (z_{10}-1) \; (z_{10}+c) \; z_{11}; \\
h_{z_{00} z_{10}} & =  -(1+c) \;  (z_{01}-1) \; (z_{01}+c) \; z_{11}; \\
h_{z_{00} z_{11}} & =  - (z_{01}-1)\; (z_{01}+c)\; (z_{10}-1) \; (z_{10}+c);\\
h_{z_{01} z_{10}} & =  -(1+c)^2 \; z_{00} \; z_{11}; \\
h_{z_{01} z_{11}} & =  -(1+c) \;  (z_{10}-1) \; (z_{10}+c) \; z_{00}; \\
h_{z_{10} z_{11}} & =  (1+c) \;  (z_{01}-1) \; (z_{01}+c) \; z_{00}.
\end{flalign*}
All of the functions $h_{ij}$ are products of linear factors; this is the
case for every equation in our classification. In other words, all of these equations are `degenerate' in the sense used in~\cite{ABS2}.
Moreover
\begin{flalign*}
r_{00} & =  (1+c)^4 \; z_{00}^2; \\
r_{11} & =   (1+c)^4 \; z_{11}^2; \\
r_{10} & =  (1+c)^2 \; (1-z_{10} )^2 \; (z_{10} +c)^2; \\
r_{01} & =  (1+c)^2 \; (1-z_{01} )^2 \; (z_{01} +c)^2. 
\end{flalign*}
These are in the canonical forms, but are not in any of the cases that were classified in Theorem 2 of~\cite{ABS2}.
Hence none of the equations that we have studied are equivalent to any equation in the ABS classification.



In summary, it is feasible to look for new integrable difference equations
by searching for equations that admit `extra' conservation laws. The class that
we have studied has been particularly fruitful, although only one of the equations
(up to equivalence) seems to be unknown. A useful by-product is that one obtains a list of
conservation laws, most of which are new (even for the known equations). The calculation
of algebraic entropy is a clear indicator of integrability and linearizability.

It is worth noting that our new equation (\ref{thtwo}) is the only one with maximal asymmetry within the form of Ansatz
(\ref{First}), because $\epsilon_1 = - \epsilon_2$ in this case alone.

Two particularly important questions remain: does the new equation have a Lax pair description,
and is it 3D-consistent?  If we wanted to check directly the
consistency around the cube, we should first choose an Ansatz for the
form of the relations we want to use on the six faces of a cube. This
leads one to ask which deformations of our models will be integrable. These might be M\"{o}bius transformations
or other deformations which do not lie within the assumed Ansatz (\ref{First}). The
analysis of the singularity pattern may be a way to tackle this
problem. One should be prepared to accept deformed equations that
are not affine; however, this is beyond the scope of the current paper.

\bigskip

\section*{Acknowledgements}

This work was carried out while the authors were visiting the Isaac Newton
Institute for Mathematical Sciences, Cambridge, UK. We thank the organisers
of the programme \textit{Discrete Integrable Systems} (January--June 2009)
for the opportunity to participate. We particularly thank Frank Nijhoff,
Kenichi Maruno, Reinout Quispel and James Atkinson for their helpful comments and
Chris Hydon for assistance with Figure 1. We are also grateful to an eagle-eyed referee.

\end{document}